\documentclass{iopart}
\usepackage{amssymb,latexsym,epsfig}
\newcommand{\beq}{\begin{equation}}
\newcommand{\eeq}{\end{equation}}
\newcommand{\bqa}{\begin{eqnarray}}
\newcommand{\eqa}{\end{eqnarray}}
\newcommand{\fr}{\frac}

\newcommand{\bra}{\langle}
\newcommand{\ket}{\rangle}

\begin{document}
\title{No-horizon theorem for spacetimes with spacelike $G_{1}$ isometry groups}
\author{S\' ergio M. C. V. Gon\c calves}
\address{Department of Physics, Yale University, New Haven, Connecticut 06511, U.S.A.}
\date{\today}
\begin{abstract}
We consider four-dimensional spacetimes $(M,{\mathbf g})$ which obey the Einstein equations ${\mathbf G}={\mathbf T}$, and admit a global spacelike $G_{1}={\mathbb R}$ isometry group. By means of dimensional reduction and local analyis on the reduced $(2+1)$ spacetime, we obtain a sufficient condition on ${\mathbf T}$ which guarantees that $(M,{\mathbf g})$ cannot contain apparent horizons. Given {\em any} $(3+1)$ spacetime with spacelike translational isometry, the no-horizon condition can be readily tested {\em without} the need for dimensional reduction. This provides thus a useful and encompassing apparent horizon test for $G_{1}$-symmetric spacetimes. We argue that this adds further evidence towards the validity of the hoop conjecture, and signals possible violations of strong cosmic censorship.
\end{abstract}
\submitto{\CQG}
\maketitle

\section{Introduction}

One of the outstanding problems in classical general relativity is that of the hoop conjecture, which puts forward a necessary and sufficient condition for horizon formation in gravitational collapse: ``Horizons form when and only when a mass $m$ gets compacted into a region whose circumference in {\em every} direction is $C\lesssim4\pi m$''~\cite{thorne72}. Despite inherent ambiguities in the definitions of horizon, mass, and circumference, no known counterexample appears to exist~\cite{nocounterhc}. In this Letter, we aim at providing strong evidence towards the ``only if'' part of the conjecture, by showing that, if mass can only be confined along {\em two} spacelike directions then apparent horizons cannot form. Our result precludes thus examples that would otherwise be blatant violations to the conjecture (e.g., infinite spindle-like configurations clothed with apparent horizons).

Thus motivated, we consider four-dimensional spacetimes $(M,{\mathbf g})$, with the minimal assumption that they admit a one-dimensional Lie group of isometries $G_{1}={\mathbb R}$ acting on a three-dimensional submanifold ${\mathcal M}$, such that $M\approx{\mathbb R}\times{\mathcal M}$. In the presence of one global spacelike Killing vector field (KVF), Einstein's equations for $(3+1)$ vacuum gravity are equivalent to $(2+1)$ gravity coupled to matter fields related to the norm and twist of the isometry-generating KVF~\cite{geroch71-72,moncrief86}. In the presence of matter, the resulting dimensionally reduced system consists of $(2+1)$ gravity coupled to an effective three-dimensional stress tensor 
\beq
^{(3)}{\mathbf T}_{\rm eff}=\,^{(3)}{\mathbf T}_{\rm g}+\,^{(3)}{\mathbf T}_{\rm m}, \label{eff}
\eeq
where
$^{(3)}{\mathbf T}_{\rm g}$ contains the four-dimensional metric's two gravitational degrees of freedom, and $^{(3)}{\mathbf T}_{\rm m}$ is the dimensionally reduced stress tensor containing the `true' (i.e., from the four-dimensional stress tensor) matter degrees of freedom. 

The two key ingredients in our analysis are: (i) the dimensional reduction of the $(3+1)$ problem to $(2+1)$ form, enabled by the global isometry, and (ii) a theorem by Ida~\cite{ida00}, which provides a sufficient condition for the absence of apparent horizons in three-dimensional spacetimes. The program is then to take a $G_{1}$-symmetric four-dimensional spacetime, perform the dimensional reduction, test for the absence of apparent horizons in the dimensionally reduced picture, and then go back to the full $(3+1)$ system and use its topological product structure to infer the topology of the apparent horizons that cannot exist therein. Natural geometrized units, in which $8\pi G=c=1$, are used throughout.

\section{Dimensional reduction}

{\bf Lemma 1.} {\em Let $(M,{\mathbf g})$ be a four-dimensional spacetime obeying the Einstein equations ${\mathbf G}[{\mathbf g}]={\mathbf T}$, and let $\xi^{\mu}$ be a globally defined spacelike Killing vector field whose space of orbits induces a three-manifold ${\mathcal M}=M/{\mathbb R}$ with Lorentzian three-metric $^{(3)}{\mathbf \theta}$. The field equations for $(M,{\mathbf g})$ are equivalent to ${\mathbf G}[^{(3)}{\mathbf \theta}]=\,^{(3)} {\mathbf T}_{\rm eff}$, where $^{(3)}{\mathbf T}_{\rm eff}$ is given by equation (\ref{eff}).} 

{\bf Proof.} We adopt Moncrief's $U(1)$ reduction approach~\cite{moncrief86}, and begin with a fully general four-dimensional spacetime $(M,\,^{(4)}g_{\mu\nu})$ and take the KVF to be $\xi^{\mu}=(\partial_{x^{3}})^{\mu}$; then $\,^{(4)}\!g_{\mu\nu}$ may be written as
\beq
ds^{2}=e^{-2\phi} \theta_{\hat{\mu}\hat{\nu}}dx^{\hat{\mu}}dx^{\hat{\nu}} + e^{2\phi}(dx^{3}+\tilde{\beta}_{a}dx^{a}+\beta_{0}dt)^{2}, \label{metric}
\eeq
where $|\xi^{\mu}|=e^{\phi}$, hatted Greek indices range over $\{0,1,2\}$, and $\theta_{\hat{\mu}\hat{\nu}}$ is the three-dimensional Lorentzian metric induced on the quotient manifold ${\mathcal M}\approx{\mathbb R}\times\Sigma$ (where $\Sigma$ is a spacelike two-surface) which admits an ADM decomposition
\beq
\theta_{\hat{\mu}\hat{\nu}}dx^{\hat{\mu}}dx^{\hat{\nu}}=-\tilde{N}^{2}dt^{2}+\tilde{\sigma}_{ab}(dx^{a}+\tilde{N}^{a}dt)(dx^{b}+\tilde{N}^{b}dt),
\eeq 
where the indices $(a,b,...)$ refer to two-dimensional quantities on $\Sigma$. One may then regard $\theta_{\hat{\mu}\hat{\nu}}$ as the natural induced metric on ${\mathcal M}$, and view $\phi$ as a smooth function, and $\beta_{0}dt+\tilde{\beta}_{a}dx^{a}$ as a smooth 1-form induced on ${\mathcal M}$. Introducing momenta $(\tilde{p}, \tilde{e}^{a}, \tilde{\pi}^{ab})$ conjugate to $(\phi, \tilde{\beta}_{a}, \tilde{\sigma}_{ab})$ in the usual way, the Einstein-Hilbert action per unit Killing length (i.e., integrated over $U=[x^{3},x^{3}+1]$ along the Killing direction) is
\beq
^{(3)}\!I=\int_{\mathcal M} dt d^{2}x (\tilde{\pi}^{ab}\tilde{\sigma}_{ab,t}+\tilde{e}^{a}\tilde{\beta}_{a,t}+\tilde{p}\phi_{,t}-\tilde{N}\tilde{\mathcal H}-\tilde{N}^{a}\tilde{\mathcal H}_{a}-\beta_{0}\tilde{e}^{a}_{,a}), \label{eha}
\eeq
where
\bqa
\tilde{\mathcal H}&=&\fr{1}{\mu_{\tilde{\sigma}}}[\tilde{\pi}^{ab}\tilde{\pi}_{ab}-(\tilde{\pi}^{a}_{a})^{2}+\fr{1}{8}\tilde{p}^{2}+\fr{1}{2}e^{-\phi}\tilde{\sigma}_{ab}\tilde{e}^{a}\tilde{e}^{b}] \nonumber \\
&-&\mu_{\tilde{\sigma}}\{^{(2)}\!R-2\tilde{\sigma}^{ab}\phi_{,a}\phi_{,b}-e^{4\phi}\tilde{\sigma}^{ac}\tilde{\sigma}^{bd}\tilde{\beta}_{[a,b]}\tilde{\beta}_{[c,d]}\}, \nonumber \\
\tilde{\mathcal H}_{a}&=&-2\tilde{\nabla}_{b}\tilde{\pi}^{b}_{a}+\tilde{p}\phi_{,a}+2\tilde{e}^{b}\tilde{\beta}_{[b,a]}, \nonumber \\
\mu_{\tilde{\sigma}}&\equiv&\sqrt{\mbox{det}(\tilde{\sigma}_{ab})}.
\eqa
The constraint equations for the action $^{(3)}\!I$ are
\beq
\tilde{\mathcal H}=0, \;\;\;\; \tilde{\mathcal H}_{a}=0, \;\;\;\; \tilde{e}^{a}_{,a}=0,
\eeq
and are equivalent to the four-dimensional constraints, restricted to the assumed symmetry class. Using the equations of motion for the canonical variables, the action (\ref{eha}) can be written as
\bqa
^{(3)}\!I&=&\int dt d^{2}x(\tilde{\pi}^{ab}\tilde{\sigma}_{ab,t}-\tilde{N}\tilde{H}-\tilde{N}^{a}\tilde{J}_{a}) \nonumber \\
&&+\int \mu_{\theta}d^{3}x \{2\phi_{,\hat{\mu}}\phi^{,\hat{\mu}}+\fr{e^{4\phi}}{2}\Psi_{\hat{\nu}}^{\;\,\hat{\mu}}\Psi^{\hat{\nu}}_{\;\hat{\mu}}\} \nonumber \\
&=&^{(3)}\!I_{\rm g}+\,^{(3)}\!I^{\rm g}_{\rm m}, \label{ham}
\eqa
where $\mu_{\theta}\equiv\sqrt{\mbox{det}(\theta_{\hat{\mu}\hat{\nu}})}$, and
\beq
\tilde{H}=\fr{1}{\mu_{\tilde{\sigma}}}[\tilde{\pi}^{ab}\tilde{\pi}_{ab}-(\tilde{\pi}^{a}_{a})^{2}]-\sqrt{\tilde{\sigma}}\,^{(2)}\!\tilde{R},\; \tilde{J}_{a}=-2\tilde{\nabla}_{b}\tilde{\pi}^{b}_{a}, \; \Psi_{\hat{\mu}\hat{\nu}}=2\beta_{[\hat{\nu},\hat{\mu}]}. \label{ha3}
\eeq
 $^{(3)}\!I_{\rm g}$ is thus just the canonical action for pure $(2+1)$ gravity, and $^{(3)}\!I^{\rm g}_{\rm m}$ is the ``matter'' action associated with the fields $\phi$ and $\tilde{\beta}_{a}$. The canonical stress tensor associated with $^{(3)}\!I^{\rm g}_{\rm m}:=\int \mu_{\theta}\,^{(3)}\!L^{\rm g}_{\rm m}$ is
\bqa
T^{\rm g}_{\hat{\mu}\hat{\nu}}&:=&-\fr{1}{\mu_{\theta}}\fr{\delta(\mu_{\theta}\,^{(3)}L^{\rm g}_{\rm m})}{\delta \theta^{\hat{\mu}\hat{\nu}}} \nonumber \\
&=&\phi_{,\hat{\mu}}\phi_{,\hat{\nu}}-\fr{1}{2}\theta_{\hat{\mu}\hat{\nu}}\phi_{,\hat{\alpha}}\phi^{,\hat{\alpha}}+\fr{e^{4\phi}}{2}(\Psi_{\hat{\mu}}^{\;\,\hat{\beta}}\Psi_{\hat{\nu}\hat{\beta}}-\fr{1}{4}\theta_{\hat{\mu}\hat{\nu}}\Psi_{\hat{\sigma}}^{\;\;\hat{\gamma}}\Psi^{\hat{\sigma}}_{\;\;\hat{\gamma}}), \label{grav}
\eqa
which has the form of a massless scalar field coupled to a source-free ``electromagnetic'' field. 

We now include a four-dimensional matter contribution. The total action per unit Killing length is
\bqa
^{(3)}I_{\rm total}&=&^{(4)}I_{\rm total}\int_{U} d\bar{x}^{3}=\int_{{\mathcal M}} \mu_{g}\, ^{(4)}\!R d^{3}x+\int_{{\mathcal M}} \mu_{g}\,^{(4)}\!L_{\rm m} d^{3}x \nonumber \\
&=& ^{(3)}I+\,^{(3)}\!I^{\rm m}_{\rm m}=\,^{(3)}\!I_{\rm g}+\,^{(3)}\!I^{\rm g}_{\rm m}+\,^{(3)}\!I^{\rm m}_{\rm m}, \label{red}
\eqa
where $^{(3)}I_{\rm g}$ and $^{(3)}I^{\rm g}_{\rm m}$ are given by equations (\ref{ham})-(\ref{ha3}), and $^{(3)}\!I^{\rm m}_{\rm m}:=\int \mu_{\theta}\,^{(3)}\!L^{\rm m}_{\rm m}$ is the dimensionally reduced `true' matter action, which is related to $^{(3)}{\mathbf T}_{\rm m}$ in equation (\ref{eff}) via
\beq
^{(3)}T^{\hat{\mu}\hat{\nu}}_{\rm m}:=\fr{1}{\mu_{\theta}}\fr{\delta(\mu_{\theta}\,^{(3)}L^{\rm m}_{\rm m})}{\delta\theta_{\hat{\mu}\hat{\nu}}}. \label{stress3}
\eeq
How does $^{(3)}T^{\hat{\mu}\hat{\nu}}_{\rm m}$ relate to $^{(4)}T^{\mu\nu}$? From equation (\ref{red}) and the definition of stress tensor in terms of a Lagrangian, we have (recalling that $^{(3)}\!I^{\rm m}_{\rm m}=\,^{(4)}\!I_{\rm m}\int_{U} d\bar{x}^{3}$)
\beq
^{(3)}\!I^{\rm m}_{\rm m}=\int \mu_{g}\,^{(4)}\!L_{\rm m} d^{3}x=\int e^{-2\phi}\mu_{\theta}\,^{(4)}T^{\mu\nu}\delta g_{\mu\nu} d^{3}x,
\eeq
and hence, using equations (\ref{metric}) and (\ref{red})-(\ref{stress3}), 
\beq
^{(3)}T^{\hat{\mu}\hat{\nu}}_{\rm m}=e^{-4\phi}\,^{(4)}T^{\hat{\mu}\hat{\nu}}_{\rm m}, \label{stress}
\eeq
where $^{(4)}T^{\hat{\mu}\hat{\nu}}_{\rm m}$ is the projection of $^{(4)}T^{\mu\nu}$ onto the submanifold orthogonal to the orbits of $\xi^{\mu}$:
\bqa
^{(4)}T^{\hat{\mu}\hat{\nu}}_{\rm m}&=&\perp^{\hat{\mu}}_{\mu}\perp^{\hat{\nu}}_{\nu}\,^{(4)}T^{\mu\nu}, \label{tproj} \\
\perp_{\mu\nu}&\equiv& g_{\mu\nu}-e^{-2\phi}\xi_{\mu}\xi_{\nu}.
\eqa

Summarizing, the four-dimensional Einstein equations $^{(4)}G_{\mu\nu}=\,^{(4)}T_{\mu\nu}$ in the presence of a global KVF of translational type are equivalent to the $(2+1)$ system $^{(3)}G_{\hat{\mu}\hat{\nu}}=\,^{(3)}T^{\rm eff}_{\hat{\mu}\hat{\nu}}\equiv\,^{(3)}T^{\rm g}_{\hat{\mu}\hat{\nu}}+\,^{(3)}T^{\rm m}_{\hat{\mu}\hat{\nu}}$, where the first term (cf. equation (\ref{grav})) contains the contributions of the four-dimensional gravitational degrees of freedom (the fields $\phi$ and $\tilde{\beta}_{a}$), and the second term (cf. equation (\ref{stress})) the true matter degrees of freedom. $\Box$

{\bf Lemma 2.} {\em The stress tensor $^{(3)}{\mathbf T}_{\rm g}$ can always be given, at least locally, in terms of wave-map fields.}

{\bf Proof.} The solution to the constraint $\tilde{e}^{a}_{,a}=0$ on the two-dimensional spacelike submanifold $\Sigma$ will depend on the latter's topology. For $\Sigma\approx{\mathbb R}^{2}$, from the Hodge decomposition of 1-forms~\cite{frankel97} it follows that one can globally define a {\em twist potential} via 
\beq
\tilde{e}^{a}:=\epsilon^{ab}\omega_{,b}, \label{twist}
\eeq
such that $\tilde{e}^{a}_{,a}\equiv0$ everywhere. If the first Betti number of $\Sigma$ is non-zero, the Hodge decomposition of $\tilde{e}_{a}$ yields an additional harmonic 1-form term, which can be made to vanish locally but not globally on $\Sigma$. However, since all of the subsequent arguments in this paper will rely solely on {\em local} analysis, the ${\mathbb R}^{2}$ topology choice for $\Sigma$ is not detrimental, and we shall henceforth adopt it. Using equation (\ref{twist}), one replaces the explicit dependence of $^{(3)}I$ on $(\tilde{\beta}^{a},\tilde{e}^{a})$ by that on $(\omega,\tilde{r})$, where $\tilde{r}\equiv(\sqrt{\tilde{\sigma}}/\tilde{N})(\omega_{,t}-\tilde{N}^{a}\omega_{,a})$ is the canonical conjugate momentum to $\omega$. In terms of these new variables, $^{(3)}\!I^{\rm g}_{\rm m}$ reads
\bqa
^{(3)}\!I^{\rm g}_{\rm m}&=&\int_{\mathcal M} dtd^{2}x\left\{\tilde{p}\phi_{,t}+\tilde{r}\omega_{,t}-\tilde{N}\left[\fr{1}{\sqrt{\tilde{\sigma}}}\left(\fr{\tilde{p}^{2}}{8}+\fr{e^{4\phi}}{2}\tilde{r}^{2}\right) \right. \right. \nonumber \\
&&\left.+\sqrt{\tilde{\sigma}}(2\tilde{\sigma}^{ab}\phi_{,a}\phi_{,b}+\fr{1}{2}e^{-4\phi}\tilde{\sigma}^{ab}\omega_{,a}\omega_{,b})\right]-\tilde{N}^{a}(\tilde{p}\phi_{,a}+\tilde{r}\omega_{,a})\} \nonumber \\
&=&\int_{\mathcal M} \mu_{\theta} d^{3}x\, \theta^{\hat{\mu}\hat{\nu}}(2\phi_{,\hat{\mu}}\phi_{,\hat{\nu}}+\fr{e^{-4\phi}}{2}\omega_{,\hat{\mu}}\omega_{,\hat{\nu}}) \nonumber \\
&=&\int_{\mathcal M} \mu_{\theta} d^{3}x\, h_{AB}\Phi^{A}_{,\hat{\mu}}\Phi^{B}_{,\hat{\nu}}\theta^{\hat{\mu}\hat{\nu}}, \label{hfaction}
\eqa
where the fields $\Phi^{A}=\phi\delta^{A}_{1}+\omega\delta^{A}_{2}$ define a mapping between the $(2+1)$ Lorentzian spacetime $({\mathcal M},\,^{(3)}\!\theta)$ and a Poincar\'{e} plane target space $V\approx{\mathbb R}^{2}$, with Riemannian metric $h_{AB}=\mbox{diag}(2,e^{-4\phi}/2)$. From Hamilton's equations, it follows that the equations of motion for $\Phi^{A}$ are just the critical points of the functional (\ref{hfaction}):
\beq
\theta^{\hat{\mu}\hat{\nu}}\nabla_{\hat{\mu}}\Phi^{A}_{,\hat{\nu}}\equiv\theta^{\hat{\mu}\hat{\nu}}(\Phi^{A}_{,\hat{\mu}\hat{\nu}}-\Gamma^{\hat{\alpha}}_{\hat{\mu}\hat{\nu}}\Phi_{,\hat{\alpha}}^{A}+\Gamma^{A}_{BC}\Phi^{B}_{,\hat{\mu}}\Phi^{C}_{,\hat{\nu}})=0, \label{hmp}
\eeq
where $\Gamma^{A}_{BC}$ are the Christoffel symbols of the metric $h_{AB}$. Equation (\ref{hmp}) is the so-called {\em wave map equation}, whose solutions are known as {\em wave maps}~\cite{hmaps}. The canonical stress tensor associated with the wave-map fields $\Phi^{A}$ is
\beq
T_{\hat{\mu}\hat{\nu}}[\Phi]:=-\fr{1}{\sqrt{|\theta|}}\fr{\delta(\mu_{\theta}\,^{(3)}T^{\rm g}_{\rm m})}{\delta \theta^{\hat{\mu}\hat{\nu}}}=\Phi_{,\hat{\mu}}\cdot\Phi_{,\hat{\nu}}-\fr{1}{2}\theta_{\hat{\mu}\hat{\nu}}\Phi_{,\hat{\alpha}}\cdot\Phi^{,\hat{\alpha}}, \label{set}
\eeq
where the dot denotes scalar product in the metric $h_{AB}$ of the target space. One can readily check that $T_{\hat{\mu}\hat{\nu}}$ is divergence-free:
\beq
\nabla_{\hat{\mu}}T^{\hat{\mu}}_{\hat{\nu}}=\Phi_{,\hat{\nu}}\cdot\theta^{\hat{\mu}\hat{\alpha}}\nabla_{\hat{\mu}}\Phi_{,\hat{\alpha}}=0,
\eeq
where the last equality follows directly from the wave map equation. $\Box$

{\bf Lemma 3.} {\em The stress-energy tensor associated with wave-map fields obeys the dominant energy condition.}

{\bf Proof.}  In what follows, $\bra,\ket$ denotes the inner product in a $(N+1)$ Lorentzian metric (with $N\geq2$), and the dot denotes the scalar product in the Riemannian metric of a target space of arbitrary dimension. The dominant energy condition (DEC) states that~\cite{wald84}, for all future-oriented timelike vector fields ${\mathbf v}$, the flux vector field ${\mathbf j}=-\bra {\mathbf T},{\mathbf v}\ket$ is future-oriented and non-spacelike. For wave-maps $\Phi$, we have (cf. equation (\ref{set})):
\beq
{\mathbf j}=-\nabla\Phi\cdot\bra{\mathbf v},\nabla\Phi\ket+\fr{1}{2}\bra\nabla\Phi\stackrel{\cdot}{,}\nabla\Phi\ket \label{jay}
\eeq
and thus
\beq
\bra{\mathbf j},{\mathbf j}\ket=\fr{1}{4}\bra{\mathbf v},{\mathbf v}\ket \bra\nabla\Phi,\nabla\Phi\ket^{2},
\eeq
which is nonpositive for $\bra{\mathbf v},{\mathbf v}\ket<0$, i.e., ${\mathbf j}$ is non-spacelike for every timelike ${\mathbf v}$. Now, since ${\mathbf v}$ is future-oriented by assumption, ${\mathbf j}$ will be too provided $\bra{\mathbf j},{\mathbf v}\ket\leq0$. From equation (\ref{jay}) this condition reads
\beq
-\bra{\mathbf v},\nabla\Phi\ket\cdot\bra{\mathbf v},\nabla\Phi\ket+\fr{1}{2}\bra{\mathbf v},{\mathbf v}\ket\bra\nabla\Phi\stackrel{\cdot}{,}\nabla\Phi\ket\leq0.
\eeq
An obvious sufficient condition for the inequality to hold is 
\beq
\bra{\mathbf v},\nabla\Phi\ket\cdot\bra{\mathbf v},\nabla\Phi\ket-\bra{\mathbf v},{\mathbf v}\ket\bra\nabla\Phi\stackrel{\cdot}{,}\nabla\Phi\ket\geq0.
\eeq
One now introduces locally Gaussian normal coordinates $\{x^{\mu}\}=\{\tau,x^{i}\}$, with $i=1...N$, wherein $\bra {\mathbf u},{\mathbf w}\ket=-u^{\tau}w^{\tau}+\Omega_{ij}u^{i}w^{j}$, and then rotate the basis vectors such that ${\mathbf v}=\partial_{\tau}$. The inequality above reads then
\beq
\Omega^{ij}\Phi_{,i}\cdot\Phi_{,j}\geq0,
\eeq
which is evidently satisfied, since both $\Omega_{ij}$ and $h_{AB}$ are Riemannian metrics. $\Box$

{\bf Lemma 4.} {\em Let $\{{\mathbf T}_{i}\}$, with $i=1...n$, be a collection of stress tensors defined on a given spacetime. If every ${\mathbf T}_{i}$ obeys the DEC, then the tensor ${\mathbf T}_{total}=\sum_{i} c_{i}{\mathbf T}_{i}$, with $c_{i}\in{\mathbb R}^{+}$, also obeys the DEC.}

{\bf Proof.} We first show that, if ${\mathbf T}_{i}$ obeys the DEC, then $c_{i}{\mathbf T}_{i}$ also does. Denoting ${\mathbf j}=-c_{i}\bra{\mathbf T}_{i},{\mathbf v}\ket$, we have
\beq
\bra {\mathbf j},{\mathbf j}\ket=c_{i}^{2}\bra\bra{\mathbf T}_{i},{\mathbf v}\ket,\bra{\mathbf T}_{i},{\mathbf v}\ket\ket <0,
\eeq
since $\bra\bra{\mathbf T}_{i},{\mathbf v}\ket,\bra{\mathbf T}_{i},{\mathbf v}\ket\ket <0$ by assumption. We also have
\beq
\bra {\mathbf j},{\mathbf v}\ket=-c_{i}\bra\bra{\mathbf T}_{i},{\mathbf v}\ket,{\mathbf v}\ket <0,
\eeq
since $\bra\bra{\mathbf T}_{i},{\mathbf v}\ket,{\mathbf v}\ket>0$ by assumption. We now show that ${\mathbf T}_{1}+{\mathbf T}_{2}$ obeys the DEC; the main result follows by induction. Denoting ${\mathbf J}=-\bra {\mathbf T}_{1}+{\mathbf T}_{2},{\mathbf v}\ket$, we have
\beq
\bra{\mathbf J},{\mathbf J}\ket=\bra\bra{\mathbf T}_{1},{\mathbf v}\ket,\bra{\mathbf T}_{1},{\mathbf v}\ket\ket + \bra\bra{\mathbf T}_{2},{\mathbf v}\ket,\bra{\mathbf T}_{2},{\mathbf v}\ket\ket+2\bra\bra{\mathbf T}_{1},{\mathbf v}\ket,\bra{\mathbf T}_{2},{\mathbf v}\ket\ket.
\eeq
The first two terms are non-positive by assumption, and the last term is also non-positive, since each ${\mathbf w}_{i}\equiv\bra{\mathbf T}_{i},{\mathbf v}\ket$ is future-oriented and non-spacelike, whereby $\bra{\mathbf w}_{1},{\mathbf w}_{2}\ket\leq0$; thus $\bra{\mathbf J},{\mathbf J}\ket\leq0$. The future-orientation of ${\mathbf J}$ follows straightforwardly:
\beq
\bra {\mathbf J},{\mathbf v}\ket=-\bra\bra{\mathbf T}_{1},{\mathbf v}\ket,{\mathbf v}\ket - \bra\bra{\mathbf T}_{2},{\mathbf v}\ket,{\mathbf v}\ket\leq0,
\eeq
since each $\bra\bra{\mathbf T}_{i},{\mathbf v}\ket,{\mathbf v}\ket\geq0$ by assumption. This completes the proof. $\Box$

{\bf Lemma 5} (Ida's Theorem). {\em Let $({\mathcal M},\,^{(3)}\!\theta)$ be a $(2+1)$ Lorentzian spacetime satisfying the Einstein equations ${\mathbf G}[^{(3)}\!\theta]=\,^{(3)}\!{\mathbf T}$. If $^{(3)}{\mathbf T}$ obeys the DEC, then there are no apparent horizons in $({\mathcal M},\,^{(3)}\!\theta)$.}

The idea of the proof consists in showing that, if an apparent horizon ${\mathcal A}$ exists {\em and} the DEC is satisfied, then one could deform ${\mathcal A}$ outward, so as to produce a new closed surface $\hat{\mathcal A}$ just outside ${\mathcal A}$, which is contained in a trapped region, thereby contradicting the ansatz that the former is the (outer marginally trapped) outer boundary of a compact trapped region. We refer the reader to Ref.~\cite{ida00,goncalves03a} for details of the proof.

\section{Absence of apparent horizons in $G_{1}$-symmetric spacetimes}

{\bf Theorem.} {\em Let $(M,\,^{(4)}g_{\mu\nu})$ be a $(3+1)$ spacetime obeying the Einstein equations $^{(4)}G_{\mu\nu}=\,^{(4)}T_{\mu\nu}$, and let $\xi^{\mu}$ be a globally defined spacelike Killing vector field whose space of orbits induces a three-manifold ${\mathcal M}=M/{\mathbb R}$. If $^{(4)}T^{\hat{\mu}\hat{\nu}}_{\rm m}$ (cf. equation (\ref{tproj})) obeys the DEC, then there are no apparent horizons in $(M,\,^{(4)}g_{\mu\nu})$.}

{\bf Proof.} Consecutive application of Lemmas 1 thru 5 implies that there are no apparent horizons in the dimensionally reduced spacetime $({\mathcal M},\,^{3}\theta)$, provided $^{(4)}T^{\hat{\mu}\hat{\nu}}_{\rm m}$ obeys the DEC. To go from the $(2+1)$ picture to the full $(3+1)$ spacetime, one must first define what one means by apparent horizon in ${\mathbb R}$-symmetric spacetimes\footnote{The translational symmetry precludes apparent horizons from being homeomorphic to $S^{2}$---the standard topology in asymptotically flat spacetimes---since one can always continuously deform any such surface along the symmetry direction (e.g., by cutting the surface along a two-plane orthogonal to the Killing direction and glueing the two parts by a topological cylinder of arbitrary length), whereby the property of outer boundary of a compact region is lost.} We shall define apparent horizons in ${\mathbb R}$-symmetric spacetimes as {\em topological $S^{1}\times{\mathbb R}$ spacelike two-surfaces which are outer marginally trapped, and are the outer boundary of a (non-compact) trapped region along the two spacelike directions on the quotient submanifold induced by the orbits of the KVF.}

The proof proceeds by {\em reductio ad absurdum}: assume that an apparent horizon exists in $(M,\,^{(4)}g_{\mu\nu})$, and then show that this implies that the reduced $(2+1)$ spacetime also contains an apparent horizon, which contradicts Ida's Theorem. Let us then assume that there is a spacelike hypersurface $^{(3)}\Sigma$ which contains the two-surface $^{(2)}{\mathcal A}\approx S^{1}\times{\mathbb R}$, which is the outer boundary of a trapped region, and satisfies 
\beq
\left[^{(4)}\nabla_{\mu}\,^{(4)}l^{\mu}\right]_{^{(2)}\!{\mathcal A}}=0, \label{omt}
\eeq 
where $^{(4)}l^{\mu}$ is the vector field tangent to future-oriented outgoing null geodesics orthogonal to $^{(2)}\!{\mathcal A}$. Now take a spacelike two-surface $^{(2)}\Sigma\approx{\mathbb R}^{2}$, such that 
$$
^{(2)}\Sigma\subset\,^{(3)}\Sigma, \;\;\;\;\;\; ^{(2)}\Sigma\,\cap\,^{(3)}\Sigma={\mathcal C}\approx S^{1},
$$ where ${\mathcal C}$ has tangent vector field $^{(3)}t^{\hat{\mu}}$ and future-oriented outgoing null normal $^{(3)}l^{\hat{\mu}}$. Consider now the $(2+1)$ spacetime obtained by the Cartesian product ${\mathbb R}\times\,^{(2)}\Sigma$, with three-metric $^{(3)}\theta_{\hat{\mu}\hat{\nu}}$. In order for ${\mathcal C}$ to be an apparent horizon in the reduced $(2+1)$ spacetime, one must have:
\bqa
^{(4)}l^{\hat{\mu}}\equiv\perp_{\mu}^{\hat{\mu}}\,^{(4)}l^{\mu}=K\,^{(3)}l^{\hat{\mu}}, \;\; K\in{\mathbb R}\backslash\{0\}, \label{co1} \\
^{(4)}l^{\hat{\mu}}\,^{(3)}t_{\hat{\mu}}=0, \label{co2} \\
^{(3)}\nabla_{\hat{\mu}}\,^{(4)}l^{\hat{\mu}}=0. \label{co3}
\eqa
Since $^{(4)}l^{\mu}$ is given by the choice for $^{(2)}{\mathcal A}$ subject to condition (\ref{omt}), and the objects $\{\phi,\beta_{0},\beta_{a},\tilde{\sigma}_{ab}\}$ are determined by the field equations, in each of the three conditions above the only free functions are (unsurprisingly) the lapse $\tilde{N}$ and shift vector $\tilde{N}^{a}$ of the three-metric $^{(3)}\theta_{\hat{\mu}\hat{\nu}}$. Conditions (\ref{co1})-(\ref{co2}) lead to a coupled system of two second-degree polynomial equations for the three variables $\{\tilde{N},\tilde{N}^{a}\}$, which one may solve for $\tilde{N}^{a}$ for a given $\tilde{N}$. Condition (\ref{co3}) is then a linear first-order PDE of gradient type for $\tilde{N}$, wherein existence and uniqueness follow from standard linear PDE theory~\cite{taylor96}. That is, given {\em any} hypersurface $^{(3)}\Sigma$ with an apparent horizon in $(M,\,^{(4)}g_{\mu\nu})$, one  can always (by appropriate gauge choice for $\theta_{\hat{\mu}\hat{\nu}}$) slice $^{(3)}\Sigma$ so as to produce an apparent horizon in the quotient $(2+1)$ spacetime. This contradicts Lemma 5 (Ida's Theorem), and thus apparent horizons cannot exist in $(M,\,^{(4)}g_{\mu\nu})$. $\Box$

The implications for the hoop conjecture are obvious: {\em any} $G_{1}$-symmetric spacetime whose dimensionally reduced stress tensor obeys the DEC is free from apparent horizons. This adds strongly to the ``only if'' part of the conjecture, by excluding cases that would otherwise be blatant violations (mass is confined along {\em two} spatial directions and yet an apparent horizon forms). The absence of large data global hyperbolicity results for the (vacuum or matter-coupled) Einstein equations with $G_{1}$ isometry allows, at least {\em a priori}~\cite{goncalves03b}, for nonspacelike singularities, which would be at least locally naked, since all nonspacelike geodesics emanating from them would be untrapped, thus in potential violation of the strong cosmic censorship conjecture.

\ack
I am grateful to Vince Moncrief for valuable discussions and comments, and to Daisuke Ida for private communications. This work was supported in part by FCT Grant SFRH-BPD-5615-2001 and by NSF Grant PHY-0098084.

\end{document}